# Excitonic Bloch-Siegert shift in CsPbI₃ perovskite quantum dots

Yuxuan Li[1,2], Yaoyao Han[1,2], Wenfei Liang[1], Boyu Zhang[1], Yulu Li,[1] Yuan Liu[1], Yupeng Yang[1], Kaifeng Wu[1]* and Jingyi Zhu[1]*.

[1] State Key Laboratory of Molecular Reaction Dynamics, Dalian Institute of Chemical Physics, Chinese Academy of Sciences, Dalian, Liaoning 116023, China.
[2] University of Chinese Academy of Sciences, Beijing 100049, China.

* Correspondence to: jingyizh@dicp.ac.cn
* Correspondence to: kwu@dicp.ac.cn

**Abstract:** Coherent interaction between matter and periodic light field induces both optical Stark effect (OSE) and Bloch-Siegert shift (BSS). Observing the BSS has been historically challenging, not only because it is weak but it is often accompanied by a much stronger OSE. Herein, by controlling the light helicity, we can largely restrict the OSE and BSS to different spin-transitions in CsPbI₃ perovskite quantum dots, achieving room-temperature BSS as strong as 4 meV with near-infrared pulses. The ratio between the BSS and OSE magnitudes is however systematically higher than the prediction by the non-interacting, quasi-particle picture. With a model that explicitly accounts for excitonic effects, we quantitatively reproduce the experimental observations. This model depicts a unified physical picture of the interplay between the OSE, biexcitonic OSE and BSS in low-dimensional materials displaying strong many-body interactions, forming the basis for the implementation of these effects to information processing, optical modulation and Floquet engineering.

**Introduction**

Quantum states of matter can be manipulated by applying intense light, examples of which include the famous optical Stark effect (OSE)[1] and Bloch-Siegert shift (BSS)[2] arising from coherent interactions between Floquet states[3] and equilibrium matter states. While originally observed in atomic/molecular systems, these effects have attracted strong attention in recent years for solid-state materials such as monolayer transition metal dichalcogenides (TMDs)[4-7] and lead halide perovskites.[8-10] Their strong light-matter interaction, in conjunction with spin-orbit coupling and/or symmetry breaking, allows for selective spin or valley state manipulation using the OSE. Moreover, the observation of a BSS, which has been historically intriguing due to its much weaker magnitude compared to the OSE, is now greatly eased thanks to the valley-exclusivity of these effects in monolayer TMDs under a quasi-particle representation.[11] Because the two time-reversed valleys $K$ and $K'$ are coupled to photons of opposite helicities, one can use helical light to pump at one valley and probe the BSS at the other one, which is naturally separated from the OSE that needs pump-probe at the same valley. By pumping with infrared pulses, a valley-exclusive BSS with its magnitude approaching that of the OSE was demonstrated.[11]

Notably, the valley-exclusivity of the BSS and OSE reported in ref 11 relies on a non-interacting quasi-particle approximation. On the other hand, however, recent studies on mono- and few-layer TMDs and many other low-dimensional material systems have in general found strong many-body interactions. These interactions induce strong excitonic and multiexcitonic effects, which will be generally called excitonic effects herein for simplicity. Under certain circumstances, the valley selection rules can be partially relaxed by excitonic effects. For example, in studies of monolayer TMDs[12,13] and CsPbBr$_3$ perovskite nanocrystals[14], with small pump detuning energies of 10s of meV, spectral modulations tunable from redshift, splitting to blueshift were observed when the pump and probe beams had opposite helicities, contradicting the simple expectation of a valley-selective OSE. This "anomalous" observation has been rationalized as a biexcitonic OSE in which the single-exciton to biexciton transition plays an important role. So far, it remains unclear how these excitonic effects influence the valley- or spin-exclusivity of the BSS.

Here we report a strong BSS up to 4 meV in CsPbI$_3$ perovskite quantum dots (QDs) at room temperature. By controlling the light helicity, we can largely restrict the OSE and BSS to different angular momentum transitions and observe them separately, especially when the pump light is tuned from visible to infrared. Importantly, the ratio between the BSS and OSE magnitudes is significantly larger than the one predicted from a non-interacting, quasi-particle picture. By diagonalizing a 9×9 Hamiltonian accounting for the co- and counter-rotating Floquet states of ground, exciton, and biexciton states, we quantitatively reproduce the experimental observations with realistic values of exciton transition dipoles and biexciton binding energies. The result also depicts a unified physical picture of the interplay between the OSE, biexcitonic OSE and BSS in materials displaying strong many-body interactions.

**Results and discussion**

**Coherent light-matter interaction.** To illustrate the fundamental principle of coherent light-matter interaction, one usually considers a simplified model of a two-level ($|g\rangle$ and $|e\rangle$) system (TLS) driven by a periodic light field, whose total Hamiltonian can be expressed as:[5,11,15]

$$\hat{H} = \hbar\omega_{eg}\hat{c}^+\hat{c} + \hbar\omega_{ph}\hat{a}^+\,\hat{a} + \hbar\lambda(\hat{c}^+\,\hat{a} + \hat{c}\,\hat{a}^+) + \hbar\lambda(\hat{c}^+\,\hat{a}^+ + \hat{c}\,\hat{a}) \qquad (1).$$

Here $\omega_{eg}$ is the transition frequency of the TLS with the creation operator $\hat{c}^+ = |\text{e}\rangle\langle\text{g}|$ and the annihilation operator $\hat{c} = |\text{g}\rangle\langle\text{e}|$, $\omega_{ph}$ is the photon frequency of the light field with similarly-defined operators $\hat{a}^+$ and $\hat{a}$, and $\lambda$ is proportional to the Rabi coupling strength containing the TLS transition dipole moment and the light field amplitude. The first and second terms in eq 1 correspond to the TLS and the photon reservoir, respectively. The third term is an energy conserving interaction corresponding to the OSE, whereas the last term is an energy non-conserving one related to the BSS. Note only one-photon interactions are considered in eq 1; the complete form of eq 1 should in principle consider a ladder of all the evenly-spaced Floquet states,[3] but higher-order terms are typically too weak compared with the one-photon terms.

When $\omega_{ph}$ is only slightly lower than $\omega_{eg}$, i.e., the energy detuning of the driving photon ($\Delta = \hbar\omega_{eg} - \hbar\omega_{ph}$) is small, the last term in eq 1 is much weaker than the third term. This corresponds to the common situation that the magnitude of the BSS is much smaller than that of the OSE. By dropping off the last term in eq 1, one obtains the original formulation of the Jaynes-Cummings model,[16] which is also called a rotating wave approximation, and the solution of this model predicts the OSE shifts reasonably well for recent experiments.[4-6,8-10] When $\omega_{ph}$ approaches zero, which can be experimentally realized by tuning the pump into infrared, the last term in eq 1 becomes important. However, by diagonalizing the full Hamiltonian in eq 1, one always obtains an energy shift as a combined result of both the OSE and BSS. Therefore, observation of an exclusive BSS, i.e., realization of the counter-rotating wave approximation, has remained intriguing for atomic/molecular and traditional solid-state systems. As mentioned in the introduction, the valley-selective transition rules in monolayer TMDs offered a unique opportunity to address this issue by separating the OSE and BSS to different valleys.[11]

**Spin-selectivity in lead halide perovskites.** Here we choose lead halide perovskites (Fig. 1a) as the material platform for the study of BSS, because their electronic structure allows us to spin-selectively address the band-edge states using helical lights, in analogy to the valley-selectivity in monolayer TMDs. Briefly, the Bloch functions of the conduction band (CB) and valence band (VB) edges have *p*- and *s*-type symmetries,[17-19] respectively, which are exactly opposite to traditional II-VI and III-V group semiconductors. Therefore, there is a large spin-orbit splitting in the CB but not in the VB. The VB edge hole states have relatively pure spin states of $m_h = \pm 1/2$, whereas the CB edge spin-orbit split-off states are $m_e = \pm 1/2$, resulting in two optically-allowed transitions coupled to photons of opposite helicities (Fig. 1b).

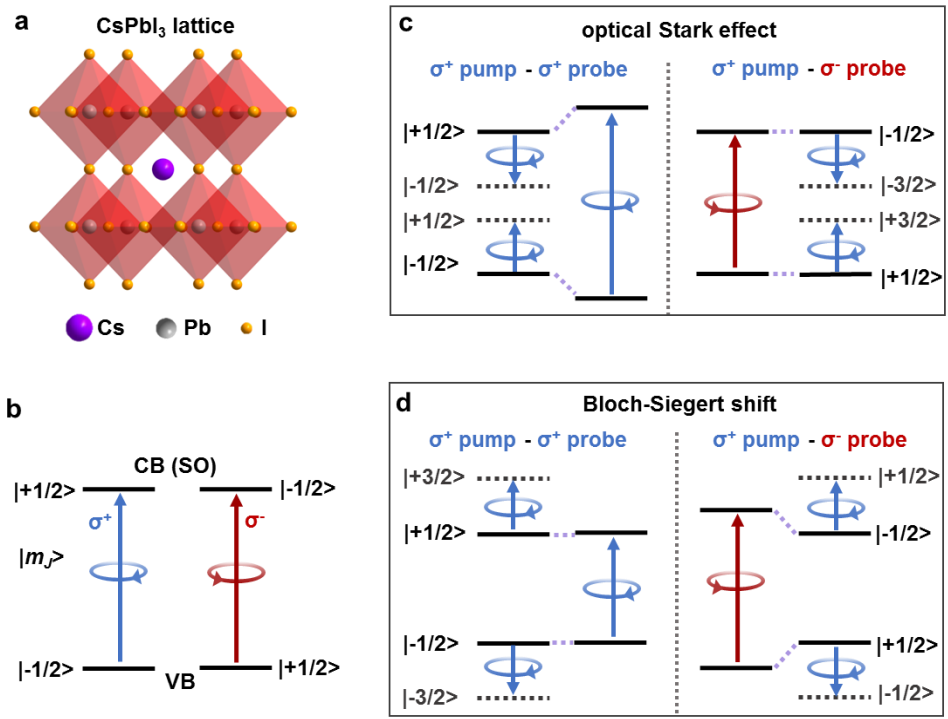

**Figure 1. Spin-selective transition rules and separation of the OSE and BSS in lead halide perovskites.** (a) Schematic illustration of the crystal structure of CsPbI$_3$

perovskite phase. (b) Conduction band (CB) and valence band (VB) edge states coupled to circularly-polarized photons. (c) Interaction energy diagram of the optical Stark shift driven by a nonresonant $\sigma^+$ photon, which occurs at the VB $|-1/2\rangle$ and CB $|+1/2\rangle$ states. The shift can be detected only with $\sigma^+\sigma^+$ pump-probe configuration. (d) Interaction energy diagram for the Bloch-Siegert shift driven by a nonresonant $\sigma^+$ photon, which occurs at the VB $|+1/2\rangle$ and CB $|-1/2\rangle$ states and can be detected only with $\sigma^+\sigma^-$ pump-probe configuration.

With this spin-selection rule and under a quasi-particle representation, one should be able to observe the spin-exclusive OSE and BSS in CsPbI$_3$ perovskites, which are depicted in Fig. 1c and 1d, respectively. By coherently driving the system with a sub-bandgap photon of left-handed helicity ($\sigma^+$), for example, the intragap Floquet states only interact with $m_h = -1/2$ and $m_e = +1/2$ states, and thus the normal OSE blueshift can be observed only with a $\sigma^+$ probe photon (Fig. 1c), as have been reported in previous studies.[8-10] In contrast, the Floquet states outside the gap only interact with $m_h = +1/2$ and $m_e = -1/2$ states, resulting in the BSS observable only with a $\sigma^-$ photon (Fig. 1d), which, however, has not been demonstrated yet.

We use strongly-confined CsPbI$_3$ QDs instead of bulk materials because the quantum confinement effect separates the band edge exciton from higher transitions, thereby further satisfying the approximation as a TLS. Moreover, the excitonic effects are strongly enhanced in QDs compared to bulk. Also importantly, these solution-processed QDs are uniformly dispersed in low-refractive-index solvents (here hexane), avoiding the dielectric disorder experienced by monolayer TMDs lying on substrates.[20] The sensitivity of the excitonic effects of monolayer TMDs to the substrates might be the reason why they are crucial in some studies but not as important in others. Thus, the perovskite QDs here are tailored to be an ideal platform to observe the BSS and to study excitonic effects on it.

**Characterization of CsPbI$_3$ perovskite QDs.** The CsPbI$_3$ QDs were synthesized by a hot-injection method.[9,21-23] A typical transmission electron microscope (TEM) image of the CsPbI$_3$ QDs is presented in Fig. 2a, displaying cuboid-shaped QDs with an average length of ~5.4 nm. The steady-state absorption spectrum of a hexane solution of 5.4 nm QDs (Fig. 2b) reveals a sharp transition located at 1.98 eV from the ground-state to the single-exciton state. This is considerably blueshifted compared to the bandgap of ~1.8 eV for bulk CsPbI$_3$ perovskites[24], a manifestation of the quantum confinement effect. Pump-probe transient absorption (TA) spectroscopy was applied to study the light-matter interactions; see details in the Methods. Fig. 2c shows the TA spectra of the CsPbI$_3$ QDs acquired using linearly-polarized pump and probe beams, at pump-probe delays of a few picoseconds following an above-gap excitation ($\hbar\omega_{ph}$ = 2.64 eV). Under this condition, real exciton populations are created, inducing state-filling and Coulombic effect signals[25-28] much stronger than the OSE and BSS. These bleach and absorption signals are situated near the exciton energy, whereas there are negligible signals ranging from ~0.7-1.7 eV as measured by our broad-band probe.

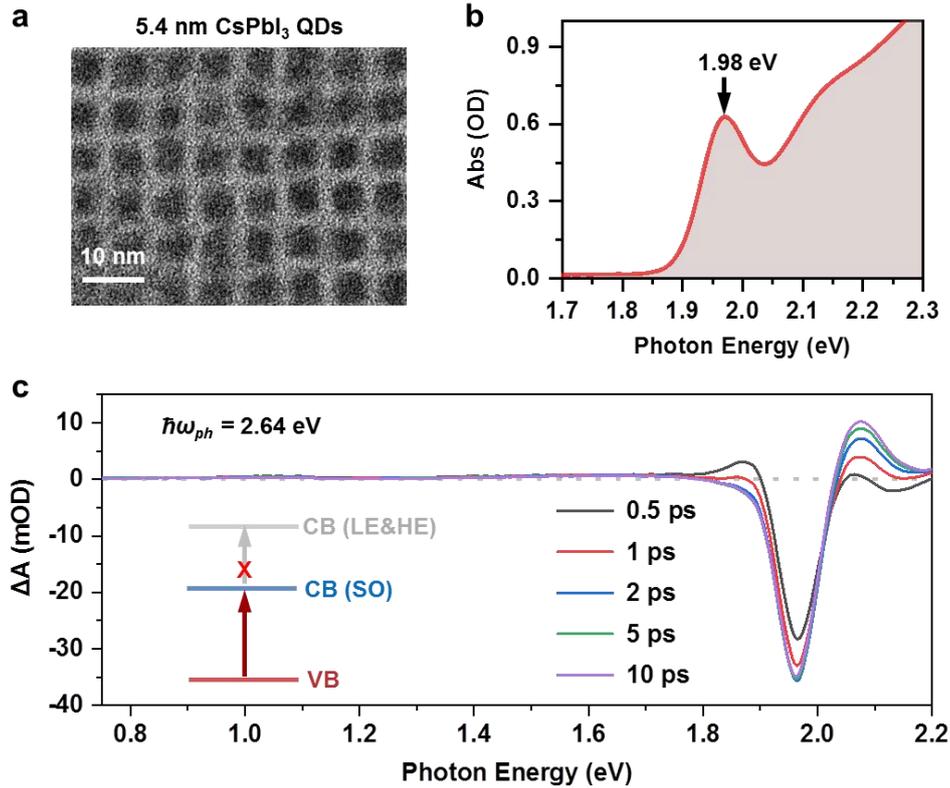

**Figure 2. Characterization of CsPbI₃ QDs.** (a) A representative transmission electron microscopy (TEM) image. (b) Steady state absorption spectrum, with the band-edge exciton peak at ~1.98 eV. (c) Broadband transient absorption (TA) spectra at varying delay times, pumped at 2.64 eV (470 nm), displaying negligible photoinduced absorption in the near infrared region. Inset is a scheme showing the forbidden nature of the intra-CB transition from band-edge spin-orbit split-off states to higher-energy light and heavy electron states.

In a recent study on CsPbBr₃ nanocrystals, a three-level system was proposed to account for anomalous Stark shifts,[29] in which an additional transition (at ~0.8 eV) from band edge exciton state to a high-energy exciton state associated with the high-energy spin-orbit electron levels in the CB was posited. The transition dipole moment was estimated to be as large as 25 Debye. However, our direct measurement using a broad-band probe clearly rules out such a transition at least in our CsPbI₃ QDs (Fig. 2c) and CsPbBr₃ QDs (Fig. S1), which is also consistent with the forbidden nature of this intraband transition (Fig. 2c inset). This allows us to limit our discussions to the band-edge exciton states only. The high-energy excitation at 2.64 eV in Fig. 2c results in dynamic effects such as hot carrier cooling and band renormalization, which are not the focus here but can be found in previous studies[30-32].

**Observation of BSS in CsPbI$_3$ perovskite QDs.** To measure the OSE and BSS, we performed pump-probe experiments using circularly-polarized beams with sub-bandgap pump photon energies ($\hbar\omega_{ph} < \hbar\omega_{eg}$). Presented in Fig. 3a, 3b and 3c are the representative TA spectra of the CsPbI$_3$ QDs with $\hbar\omega_{ph}$ of 1.74 eV (714 nm; 0.30 GW/cm$^2$), 1.21 eV (1027 nm; 1.71 GW/cm$^2$) and 0.805 eV (1540 nm; 1.86 GW/cm$^2$), respectively, at time-zero. Note that because the pump intensity varies, we focus on the relative signal size of $\sigma^+\sigma^+$ and $\sigma^+\sigma^-$ pump-probe configurations in these plots. In contrast to Fig. 2c where it shows the real exciton population with negligible decay on the ps timescale, the signals here are not associated with any real population excitation through multiphoton absorption, as they dominate only at around time-zero when the pump and probe beams temporally overlap (Fig. 3d, 3e and 3f). For $\hbar\omega_{ph} =$ 1.74 eV (Fig. 3a), the detuning is relatively small ($\Delta = 0.24$ eV), and a strong spectral modulation corresponding to the normal OSE blueshift is detected with a $\sigma^+\sigma^+$ pump-probe configuration. In contrast, the $\sigma^+\sigma^-$ configuration reveals a very weak spectral modulation. By decreasing $\hbar\omega_{ph}$, the $\sigma^+\sigma^-$ spectral modulation progressively grows until it becomes comparable with $\sigma^+\sigma^+$ at $\hbar\omega_{ph} = 0.805$ eV (Fig. 3c).

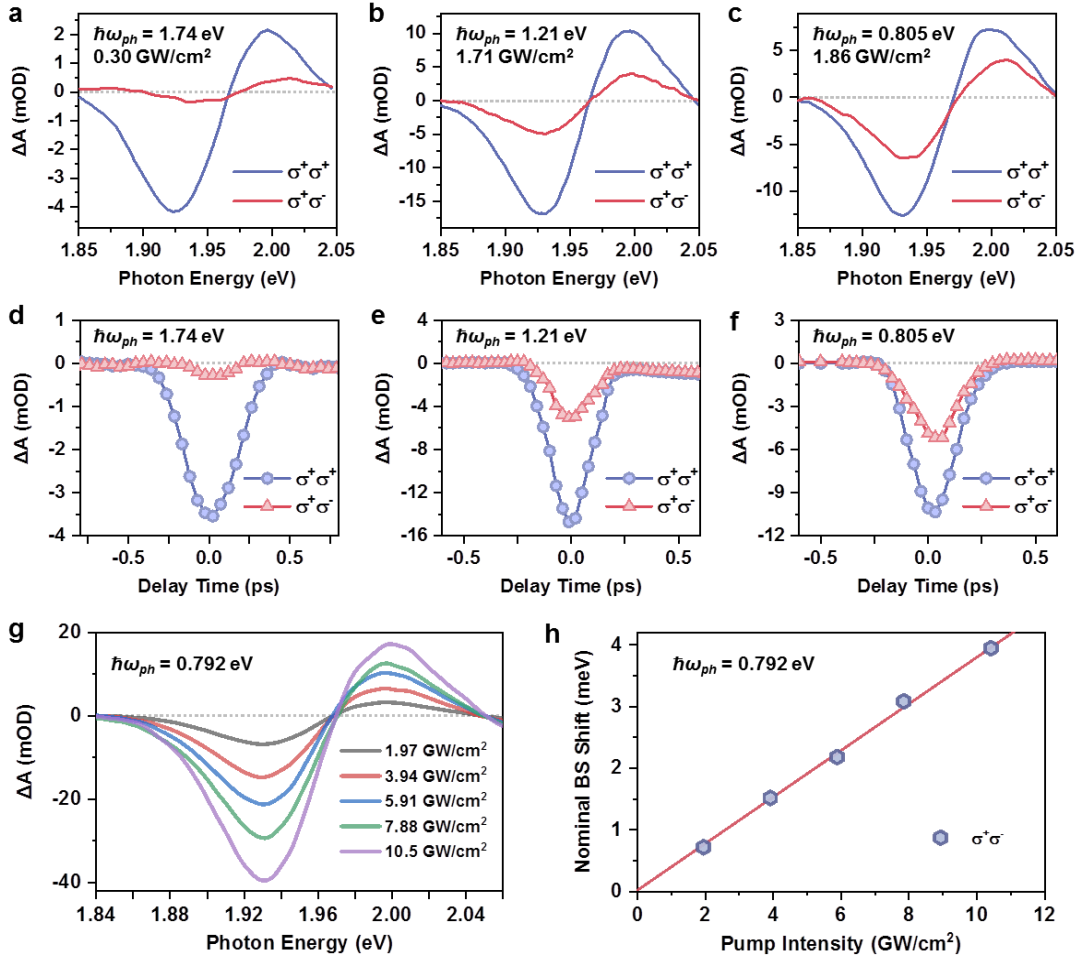

**Figure 3. Optical Stark (σ⁺σ⁺) and Bloch-Siegert (σ⁺σ⁻) shifts in CsPbI₃ QDs.** (a-c) Time-zero transient absorption (TA) spectra obtained with σ⁺σ⁺ (blue) and σ⁺σ⁻ (red) pump-probe configurations with pump photon energies at (a) 1.74 eV (714 nm), (b) 1.21 eV (1027 nm), and (c) 0.805 eV (1540 nm). The incident pump intensities are indicated in each panel. (e-f) Temporal profiles of the signals reported in (a-c). Note that the 1.74 eV pump pulse has been spectrally filtered using a grating in order to avoid resonant excitation, and consequently the pulse duration is broadened. (g) σ⁺σ⁻ TA spectra under varying pump intensities of 0.792 eV (1565 nm) photon. (h) Nominal Bloch-Siegert shift (blue circles) as a function of the pump intensity, calculated from the spectral-weight-transfer in (g). The red solid line is a linear fit.

The detuning-dependent behaviors observed in Fig. 3a-3c are phenomenologically consistent with the spin-selective OSE and BSS of the CsPbI₃ QDs that we depict in Fig. 1c and 1d. By diagonalizing the Hamiltonian in eq 1 under co- and counter-rotating wave approximations, respectively, one can obtain the spectral blueshifts induced by the pure OSE and BSS as (see Supplementary Text):

$$\delta E_{OSE} = 2\hbar\lambda^2/(\omega_{eg} - \omega_{ph})$$

$$\delta E_{BSS} = 2\hbar\lambda^2/(\omega_{eg} + \omega_{ph}) \qquad (2).$$

Thus, the ratio between them is:

$$\delta E_{BSS}/\delta E_{OSE} = (\omega_{eg} - \omega_{ph})/(\omega_{eg} + \omega_{ph}) \qquad (3),$$

which should increase with a decreasing $\omega_{ph}$, as we observed herein.

In Fig. 3g, we plot the time-zero $\sigma^+\sigma^-$ TA spectra measured with $\hbar\omega_{ph} = 0.792$ eV of varying power densities, which are nominally dominated by the BSS. We quantify the spectral shift using a spectral-weight-transfer calculation[5] and obtain power-dependent shift plotted in Fig. 3h. The signal increases linearly with the pump power, further substantiating that it is driven by single-photons. Note that the pump power was measured outside the sample cuvette, which is overestimated compared to the local power inside the QD that is screened by the solvent and QD materials. At an outside pump power of ca. 10 GW/cm$^2$, the nominal BSS has reached 4 meV, which is several orders of magnitude larger than those reported for superconducting-qubit TLS[33,34] and is comparable to the shift observed in the monolayer TMDs[11]. The large nominal BSS arises from the combined contribution of the small pump photon energy and the inherently strong light-matter interaction in the CsPbI$_3$ QDs.

While the BSS and OSE signals followed eq 2 reasonably well in ref 11, in the presence of strong many-body excitonic effects the precision of this simple model needs to be reconsidered. This is the reason we call the shifts in Fig. 3 nominal BSS and OSE. To uncover the role of excitonic effects, we tune $\hbar\omega_{ph}$ to 12 different energies and measure the spectral shifts under $\sigma^+\sigma^+$ and $\sigma^+\sigma^-$ configurations. The associated TA spectra at time-zero are presented in Fig. S2. Quantitative comparison between our experimental results at different $\hbar\omega_{ph}$ with eq 2, however, is difficult (Fig. S3), because the pump power screening by the sample depends on $\hbar\omega_{ph}$ and the power density measurements for different $\hbar\omega_{ph}$ might introduce large errors. Nevertheless, by taking the ratio between the nominal $\delta E_{BSS}$ and $\delta E_{OSE}$, these $\hbar\omega_{ph}$-dependent issues can be cancelled, allowing us to compare our experimental results to eq 3. Fig. 4a shows the plot of the ratios at 12 different $\hbar\omega_{ph}$, the general trend of which is in line with eq 3. However, the experimental ratio is systematically larger than the one predicted by eq 3, and the deviation is $\hbar\omega_{ph}$-dependent. At small detuning (e.g., $\hbar\omega_{ph}$ ~1.8 eV), the deviation is small, but it becomes large as the detuning increases.

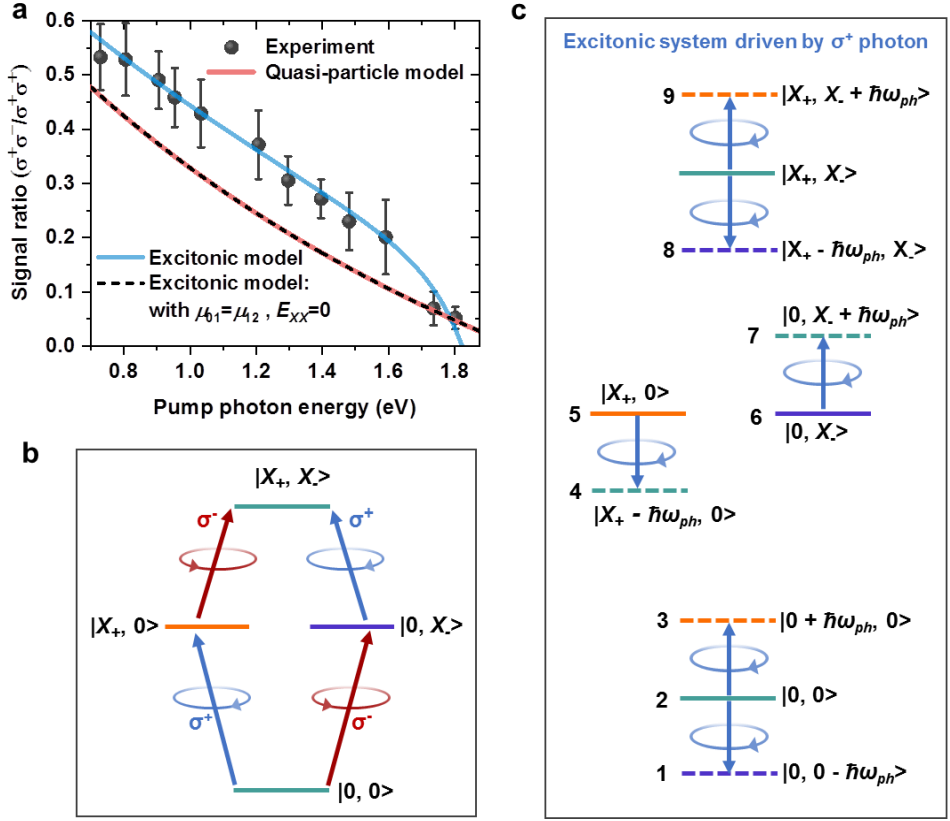

**Figure 4. Excitonic Bloch-Siegert Shift in CsPbI$_3$ QDs.** (a) Ratio between the normal Bloch-Siegert and optical Stark shifts measured with $\sigma^+\sigma^-$ and $\sigma^+\sigma^+$ configurations, respectively (gray circles). The red solid line is the prediction by a non-interacting, quasi-particle model (eq 3), whereas the blue solid line is obtained using our excitonic model (eq 4) with $\mu_{01} = 34$ D, $\mu_{12} = 28$ D and $E_{XX} = 65$ meV. By setting $\mu_{01} = \mu_{12}$ and $E_{XX} = 0$, our excitonic model (black dashed line) reduces to the quasi-particle model. The error bars are obtained from the standard variations of $\sigma^+\sigma^+$ and $\sigma^+\sigma^-$ signals in the range of ±40 fs around time-zero. (b) States and optical selection rules under the excitonic representation. (c) Equilibrium matter states (solid lines) and Floquet states (dashed lines) driven by a nonresonant $\sigma^+$ photon, whose angular momenta are indicated by their colors. Other Floquet states are not shown because their interactions with matter states are against angular momentum conservation. The shift of the biexciton state $|X_+, X_-\rangle$ is not considered here as we are experimentally probing the shift of single-exciton transitions. There are 9 states in total in the interaction Hamiltonian.

We note that a deviation from the cubic symmetry by either an orthorhombic lattice distortion or a QD shape effect can result in linearly-polarized excitons instead of circularly-polarized ones through anisotropic exchange interaction, as revealed in recent single-particle photoluminescence studies on perovskite QDs at liquid-helium temperatures.[17,35,36] In this case, $\sigma^+\sigma^+$ and $\sigma^+\sigma^-$ configurations should yield the same signal size, i.e. the ratio between $\sigma^+\sigma^+$ and $\sigma^+\sigma^-$ signals should be 1 regardless the $\hbar\omega_{ph}$ used. Even if there is only a small portion of QDs displaying such a behavior, the deviation of the experimental data from the prediction by eq 3 should monotonically increase with $\hbar\omega_{ph}$, which is just opposite to our results here. Therefore, circular-polarization selection rules still largely hold here.

A possible reason is that the anisotropy-induced exciton fine structure splitting is very weak, on the order of μeV-meV as determined from recent single-dot[17,35-37] and ensemble[38] measurements on lead halide perovskite QDs. This splitting is much smaller in magnitude compared to the single-dot exciton linewidth of 50-100 meV for these QDs.[39,40] In this limit, it might still be reasonable to use the angular momentum basis of quasi-cubic approximation in Fig. 1. Indeed, when we perform linear-polarization OSE measurements for our 5.4 nm $CsPbI_3$ QDs, we find that the co- and cross-linear configurations give almost identical signal sizes (Fig. S4). This observation is inconsistent with ideal linear-polarization selection rules which predict a signal ratio of 3:1 for co- over cross-linear configurations under orientational averaging. It is instructive to compare these results to a previous study on few-layer ReS$_2$ flakes.[41] Therein, the anisotropy-induced exciton splitting reaches 60 meV, much larger than the room-temperature exciton linewidth of ~25 meV for this material, resulting in clean linearly-polarized OSE features.[41] This comparison lends further support to our linewidth-versus-splitting hypothesis.

**Excitonic BSS model.** To explain the observed deviation, we invoke a model taking into consideration the excitonic effects. Under an excitonic representation, the optical selection rules between the ground state $|0, 0\rangle$, the single-exciton states $|X_+, 0\rangle$ and $|0, X_-\rangle$, and the biexciton state $|X_+, X_-\rangle$ are depicted in Fig. 4b. When the system is coherently driven by a detuned $\sigma^+$ photon, the six one-photon-dressed Floquet states that have strong interactions with the three equilibrium states of the ground state $|0, 0\rangle$ and the single-exciton states $|X_+, 0\rangle$ and $|0, X_-\rangle$ are considered, as shown in Fig. 4c. Other Floquet states whose interactions with the single-exciton states $|X_+, 0\rangle$ and $|0, X_-\rangle$ are against angular momentum conservation are not included here.[11] The hybridizations between the states in Fig. 4c can be analyzed by diagonalizing the 9×9 Hamiltonian:

$$\hat{H} = \begin{pmatrix} -\hbar\omega & 0 & 0 & 0 & 0 & \mu_{01}\varepsilon & 0 & 0 & 0 \\ 0 & 0 & 0 & \mu_{01}\varepsilon & 0 & 0 & \mu_{01}\varepsilon & 0 & 0 \\ 0 & 0 & \hbar\omega & 0 & \mu_{01}\varepsilon & 0 & 0 & 0 & 0 \\ 0 & \mu_{01}\varepsilon & 0 & E_1-\hbar\omega & 0 & 0 & 0 & 0 & 0 \\ 0 & 0 & \mu_{01}\varepsilon & 0 & E_1 & 0 & 0 & 0 & \mu_{12}\varepsilon \\ \mu_{01}\varepsilon & 0 & 0 & 0 & 0 & E_1 & 0 & \mu_{12}\varepsilon & 0 \\ 0 & \mu_{01}\varepsilon & 0 & 0 & 0 & 0 & E_1+\hbar\omega & 0 & 0 \\ 0 & 0 & 0 & 0 & 0 & \mu_{12}\varepsilon & 0 & E_2-\hbar\omega & 0 \\ 0 & 0 & 0 & 0 & \mu_{12}\varepsilon & 0 & 0 & 0 & E_2+\hbar\omega \end{pmatrix} \quad (4).$$

The diagonal elements are arranged in the order of numbered states (from 1 to 9) as indicated in Fig. 4c, with the energy of $|0, 0\rangle$ set at 0, and the energies of $|X_+, 0\rangle$ ($|0, X_-\rangle$) and $|X_+, X_-\rangle$ at $E_1$ and $E_2$, respectively. $E_1$ and $E_2$ are related by: $E_2 = 2E_1 - E_{XX}$, where $E_{XX}$ is the biexciton interaction energy (positive for attraction). The transition dipole moment of $|0, 0\rangle$ to $|X_+, 0\rangle$ ($|0, X_-\rangle$) is $\mu_{01,}$ and $|X_+, 0\rangle$ ($|0, X_-\rangle$) to $|X_+, X_-\rangle$ is $\mu_{12}$, which are coupled to the driving field $\varepsilon$.

By diagonalizing eq 4, we can well reproduce the experimental ratio of the nominal $\delta E_{BSS}$ and $\delta E_{OSE}$, as compared in Fig. 4a. In the simulation, we use $E_{XX} = 65$ meV, $\mu_{01} = 34$ D, and $\mu_{12} = 28$ D. The biexciton interaction energy results from a detailed competition between the repulsive and attractive energy parts of Coulomb and exchange interactions, which is generally attractive in low-dimensional materials,[42,43] unless unconventional electron and hole wavefunctions were deliberately designed[44]. The single-exciton transition dipole ($\mu_{01}$) of 34 D is similar to the one determined from a previous OSE study of similar-size CsPbI$_3$ QDs.[9] The reason for a reduced magnitude of $\mu_{12}$ than $\mu_{01}$ is likely associated with the dielectric contrast effect for the inorganic QDs surrounded by low-refractive-index organic ligands and solvent, which enhances $\mu_{01}$ more significantly than $\mu_{12}$, as predicted by a previous calculation[45].

To test the generality of our model, we measured another two samples of CsPbI$_3$ QDs with average edge lengths of ~4.0 nm and 7.4 nm. The relevant data are presented in Fig. S5. The ratio between signals obtained with $\sigma^+\sigma^-$ and $\sigma^+\sigma^+$ configurations ($\sigma^+\sigma^-/\sigma^+\sigma^+$) for these two samples also deviate from the non-interacting, quasi-particle model, and can be well reproduced by our excitonic model. From the model fits, we obtain $\mu_{01} = 26$ D (50 D), $\mu_{12} = 19$ D (41 D), and $E_{XX} = 90$ meV (48 meV) for 4.0 nm (7.4 nm) QDs; See Table S1. Therefore, with decreasing QD size, both $\mu_{01}$ and $\mu_{12}$ become smaller, whereas $E_{XX}$ is enhanced, which are consistent with the physical properties of QDs. We note that some recent studies reported repulsive biexciton interactions in lead halide perovskite QDs,[46-48] which are still under debate are out of the scope of our current study. By contrast, recent single-dot PL measurements at cryogenic temperatures allow to directly readout exciton and biexciton emission lines, which gave biexciton binding energies of ~20 meV for weakly-confined (~10 nm) $CsPbI_3$ QDs.[36,37] These results are in line with the extrapolation of the size-dependent trend of $E_{XX}$ obtained herein.

## Discussion

Our new model indicates that the OSE, biexcitonic OSE and BSS are partially mixed under the many-body excitonic representation. According to Fig. 4c, the key differences to the non-interacting quasi-particle picture are: i) the ground state $|0, 0\rangle$ is simultaneously repelled down by $|X_+ - \hbar\omega_{ph}, 0\rangle$ and $|0, X_- + \hbar\omega_{ph}\rangle$ through the OSE and BSS, respectively; ii) the single-exciton state $|0, X_{-1}\rangle$ state is repelled down by $|X_+ - \hbar\omega_{ph}, X_-\rangle$ through the biexcitonic OSE but repelled up by $|0, 0 - \hbar\omega_{ph}\rangle$ through the BSS; and iii) the single-exciton state $|X_{+1}, 0\rangle$ state is repelled up by $|0 + \hbar\omega_{ph}, 0\rangle$ through the OSE but also repelled down by $|X_+, X_- + \hbar\omega_{ph}\rangle$ through the biexcitonic BBS. In the scenario of unbound biexcitons ($\mu_{01} = \mu_{12}$ and $E_{XX} = 0$), the first shift terms in i) and ii) are exactly the same, as are the second shift terms in i) and iii), and hence, the excitonic model naturally reduces to the quasi-particle model for which a spin- or valley-selective OSE or BSS can be observed. This is confirmed in our simulation in Fig. 4a. In the case of bound biexcitons, the net result of the above shifts is a generally enhanced $\sigma^+\sigma^-/\sigma^+\sigma^+$ signal ratio. Still, there exist certain driving photon energies for which the total shifts in i) and ii) can be almost identical, thus canceling the signal in $\sigma^+\sigma^-$ configuration. One situation in Fig. 4a is when $\hbar\omega_{ph}$ is around 1.8-1.9 eV, i.e., the detuning is 100-200 meV. In this case, the apparent result is that a so-called spin- or valley-selective OSE can be detected (i.e., $\sigma^+\sigma^-$ yields a negligible shift; see the spectra in Fig. 3a for example). Indeed, detuning energies in that range were typically used in previous studies of lead halide perovskites[8-10] and monolayer TMDs[4,5] focusing on the spin- and valley-selective OSE. Further tuning $\hbar\omega_{ph}$ towards the exciton energy should result in spectral shifts dominated by the biexcitonic OSE,[13,14] which however is not the focus here. Moreover, such measurements can be very challenging for our ensemble sample due to inhomogeneous broadening of the exciton linewidth, which will result in real population excitation.

To summarize, this study reports the observation of a strong Bloch-Siegert shift in CsPbI$_3$ perovskite QDs, and more importantly, it highlights the importance of many-body excitonic effects in the correct interpretation of coherent light-matter interaction even for materials with spin- or valley-selective selection rules. With these excitonic interactions, the single-exciton optical Stark effect, biexcitonic optical Stark effect and Bloch-Siegert shift are intermixed. Considering the ubiquitous role of strong excitonic effects in quantum-confined materials, the model developed here should be transferrable to many other such systems. This knowledge is crucial for the implementation of the above coherent effects to information processing and optical modulation, and to the emerging field of Floquet engineering of quantum materials[49,50]. Although steady-state Floquet engineering is highly desired,[51] these instantaneous engineering techniques in turn allow for ultrafast manipulations that are particularly useful for very high-speed quantum computing.[52]

**Methods**

The CsPbI$_3$ QDs were synthesized via a hot-injection method described elsewhere.[22] Pump-probe measurements were conducted using a Pharos Yb: KGW laser (Light Conversion; 1030 nm, 230 fs pulse-duration, 100 kHz repetition rate) as the laser source. The output laser was split into multiple beams, one of which was directed to an optical parametric amplifier (Orpheus-HP; Light Conversion) to generate wavelength-tunable pump pulses, and the other being delayed by a motorized delay stage and then focused onto a YAG crystal to generate a white light continuum as the probe. In order to eliminate the potential polarization distortion during propagation of beams, circular polarization of pump and probe beams was produced by separately inserting two sets of beam-splitting cubes (Thorlabs) and quarter-wave plates (Thorlabs) right before the sample. The pump and probe directions were set to a small crossing angel at ~6 °, and focused onto the sample with long focal lens of 40 and 25 cm, respectively. For all the detuned pump laser wavelengths, the sizes of the foci on the sample were measured to be in the range 250-300 μm, more than twice larger than the probe beam size (~120 μm). The transmitted probe beam was re-collimated and focused into a fiber-coupled spectrometer to record the variation in absorbance of the sample induced by pump pulses. The sample was placed into a 1 mm quartz cuvette and all measurements were carried out at room temperature (~298 K).

**Acknowledgments**

K.W. acknowledged financial support from the the Ministry of Science and Technology of China (2018YFA0208703), the Chinese Academy of Sciences (YSBR-007), and Dalian Institute of Chemical Physics (DICP I201914).

**Author contributions**

K.W. supervised the study and designed the project. Y.H., Y.L. and W.L. synthesized the samples. Y.L. and J.Z. measured their spectroscopy. J.Z. developed the Hamiltonian for the simulation. B.Z., Y.L., Y.L. and Y.Y. helped with sample synthesis or spectroscopy experiments. K.W. and J.Z. wrote the manuscript with contributions from all authors.

**Competing interests:** Authors declare no competing interests.

**Data and materials availability:** All data is available in the main text or the supplementary materials and can be obtained upon reasonable request from K.W. (kwu@dicp.ac.cn).

**Code availability:** Custom software developed for theoretical modeling associated with this study is available for verification purposes upon reasonable request from J.Z (jingyizh@dicp.ac.cn).

**Supplementary Information:** Available online

# Excitonic Bloch-Siegert shift in CsPbI$_3$ perovskite quantum dots

Yuxuan Li[1,2], Yaoyao Han[1,2], Wenfei Liang[1], Boyu Zhang[1], Yulu Li,[1] Yuan Liu[1], Yupeng Yang[1], Kaifeng Wu[1]* and Jingyi Zhu[1]*.

[1] State Key Laboratory of Molecular Reaction Dynamics, Dalian Institute of Chemical Physics, Chinese Academy of Sciences, Dalian, Liaoning 116023, China.
[2] University of Chinese Academy of Sciences, Beijing 100049, China.

* Correspondence to: jingyizh@dicp.ac.cn
* Correspondence to: kwu@dicp.ac.cn

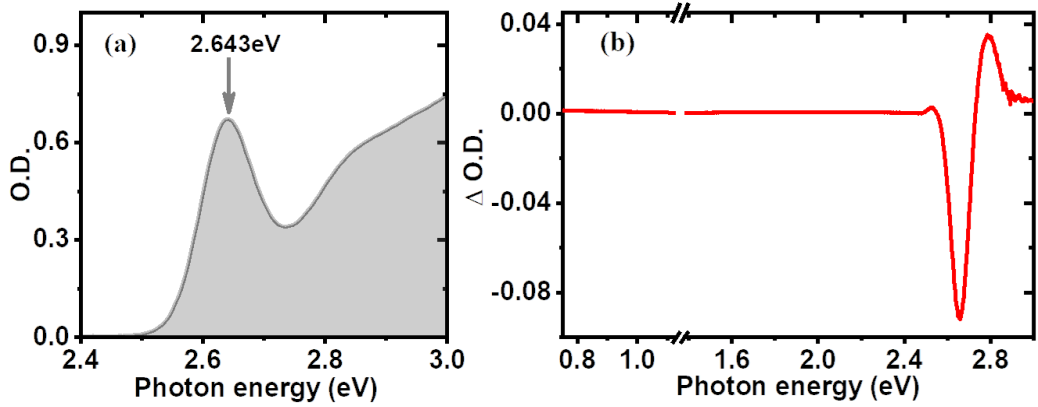

**Figure S1. Spectra of CsPbBr₃ QDs.** (a) Steady state absorption spectrum and (b) transient absorption spectrum at 2 ps of the 3.9 nm CsPbBr₃ QDs. Pump photon energy is at around 3.02 eV (410 nm) and is linearly polarized.

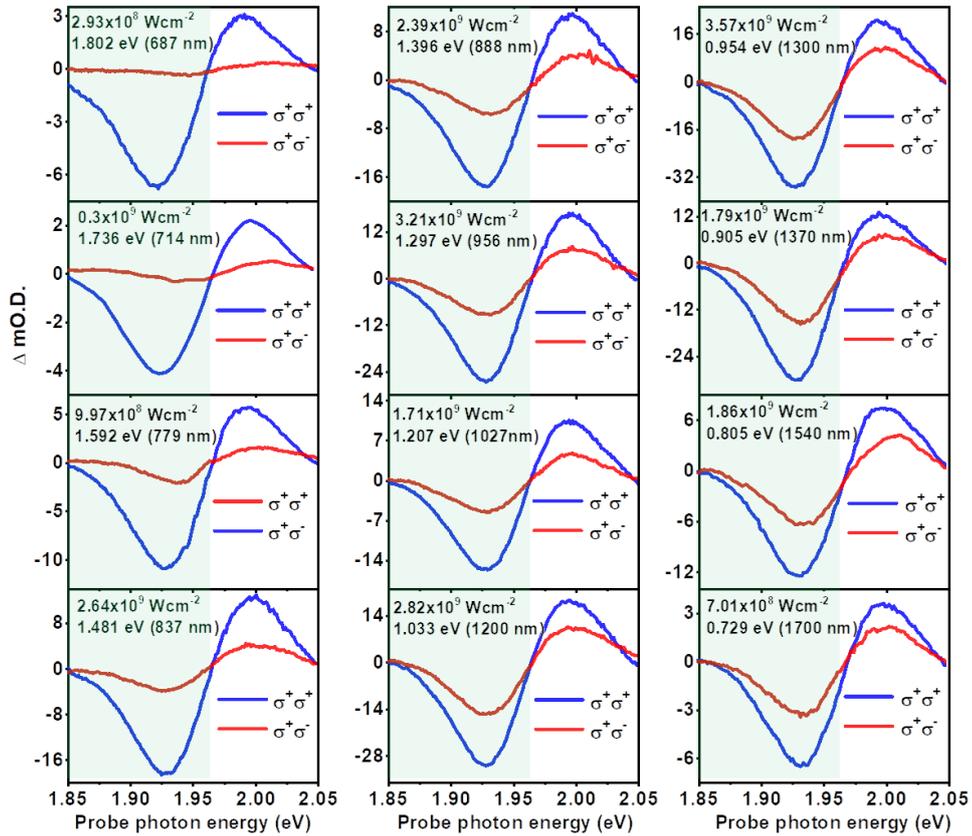

**Figure S2. TA spectra.** Pump photon energy dependence of the nominal Stark ($\sigma^+\sigma^+$) and Bolch-Siegert ($\sigma^+\sigma^-$) shift TA spectra observed in CsPbI₃ quantum dots. The pump photon energies and intensities are indicated in each panel, and the shallow green shaded regions are used for the spectral- weight-transfer calculations for obtain the ratio of $\sigma^+\sigma^-/\sigma^+\sigma^+$ in the main manuscript.

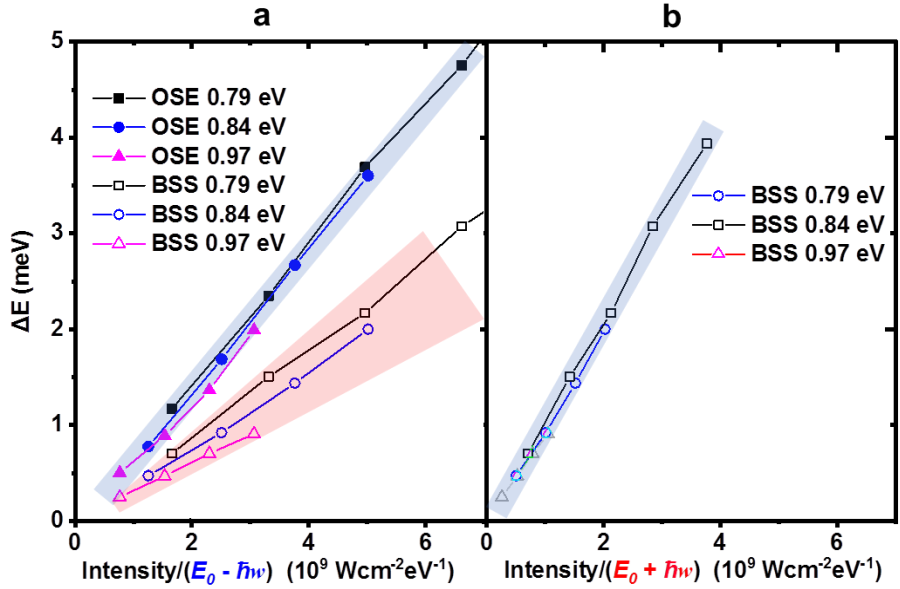

**Figure S3. Plots of $\hbar\omega_{ph}$-dependent OSE and BSS signals.** (a) Plots of nominal energy shifts induced by OSE and BSS (measured with σ⁺σ⁺ and σ⁺σ⁻ configurations, respectively) as a function of pump intensity over ($E_0 - \hbar\omega_{ph}$), where $E_0$ is the exciton energy. Three different pump photon energies $\hbar\omega_{ph}$ are included in the plot. (b) Re-plot of BSS signals as a function of pump intensity over ($E_0 + \hbar\omega_{ph}$).

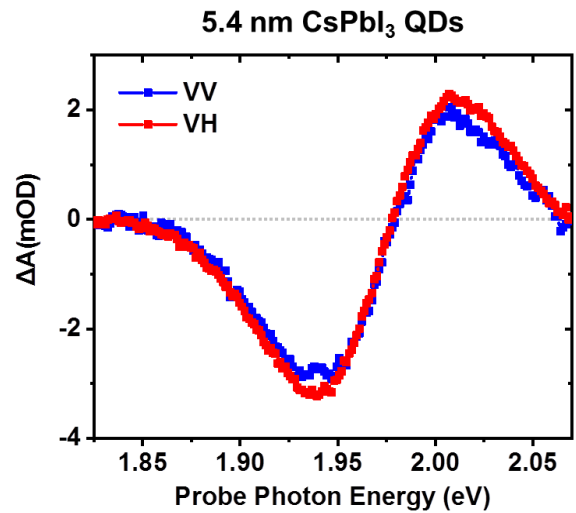

**Figure S4. Linear-polarization OSE.** Linear-polarization pump-probe (VV and VH) measurements of the OSE in 5.4 nm CsPbI$_3$ QDs ($\hbar\omega_{ph}$: 1.72 eV; 0.4 GW/cm$^2$ ).

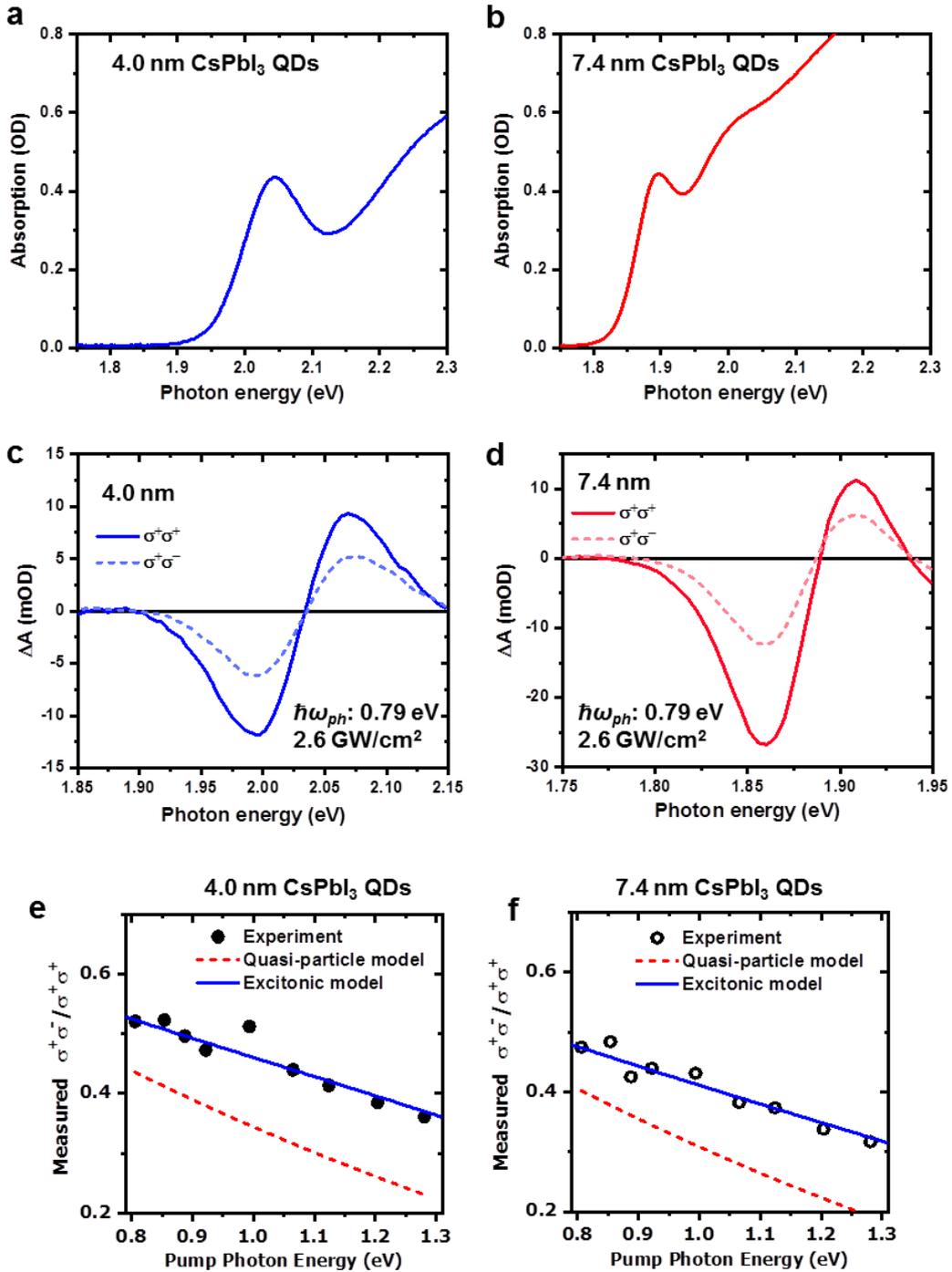

**Figure S5. QD size-dependent measurements.** (a,b) Absorption spectra of another two sets of CsPbI$_3$ QDs, with their average edge lengths of (a) 4.0 nm and (b) 7.4 nm. (c,d) TA spectra measured with σ⁺σ⁺ (solid) and σ⁺σ⁻ (dashed) configurations for (c) 4.0 nm and (d) 7.4 nm QDs. The pump conditions are indicated. (e,f) Ratio between signals obtained with σ⁺σ⁻ and σ⁺σ⁺ configurations (σ⁺σ⁻/σ⁺σ⁺) for (e) 4.0 nm and (f) 7.4 nm QDs. The experimental data are in black circles, fittings using the quasi-particle and excitonic models are shown as red dashed and blue solid lines, respectively.

**Table S1. Summary of the $\mu_{01}$, $\mu_{12}$ and $E_{XX}$ parameters for QDs of three sizes.**

|         | $\mu_{01}$ (Debye) | $\mu_{12}$ (Debye) | $E_{XX}$ (meV) |
|---------|--------------------|--------------------|----------------|
| **4.0 nm** | 26 | 19 | 90 |
| **5.4 nm** | 34 | 28 | 65 |
| **7.4 nm** | 50 | 41 | 48 |

**Supplementary Text: Optical Stark and Bloch-Siegert shift of a two-energy level system.**

For a simple two-energy level system (ground state *G* and excited state *Ex*) with transition dipole moment of $\mu_{eg}$ and resonant energy of $\hbar\omega_{eg}$, we can resolve two situations of the interaction with the detuned light field: the corotating interaction component and the counterrotating interaction, corresponding to the optical Stark and Bloch-Siegert shifts, respectively. Following eq 1 in the main manuscript, for the corotating part, the Hamiltonian can be written as:

$$\hat{H}_{Stark} = \hbar\omega_{eg}\hat{c}^{\dagger}\hat{c}^{-} + \hbar\omega_{ph}\hat{a}^{\dagger}\hat{a}^{-} + \hbar\lambda(\hat{c}^{\dagger}\hat{a}^{-} + \hat{c}^{-}\hat{a}^{\dagger}) \dots\dots\dots (1)$$

In the subspace containing the excited matter-state $|Ex, 0\rangle$ and the one-photon dressed up Floquet state $|G, +1\rangle$, (1) becomes:

$$\hat{H}_{Stark}^{up} = \hbar\omega_{eg}\begin{bmatrix} 1 & 0 \\ 0 & 0 \end{bmatrix} + \hbar\omega_{ph}\begin{bmatrix} 0 & 0 \\ 0 & 1 \end{bmatrix} + \hbar\lambda\begin{bmatrix} 0 & 1 \\ 1 & 0 \end{bmatrix} \dots\dots\dots (2)$$

The diagonalized eigen energy is

$$E_{Stark}^{up} = \frac{\hbar}{2}(\omega_{eg} + \omega_{ph}) \pm \frac{\hbar}{2}\sqrt{(\omega_{eg} - \omega_{ph})^2 + 4\lambda^2} \ \dots\dots (3)$$

For large negative detuning and moderate coupling strength, i.e., $\lambda \ll \omega_{eg} - \omega_{ph}$, (3) reduces to:

$$E_{Stark}^{up} = \frac{\hbar}{2}(\omega_{eg} + \omega_{ph}) \pm \frac{\hbar}{2}(\omega_{eg} - \omega_{ph}) \pm \frac{\hbar\lambda^2}{(\omega_{eg} - \omega_{ph})} \ \dots\dots (4)$$

For large detuning, only the high-energy level containing the matter property can be coupled to the probe pulse, whose energy is:

$$E^{up}_{Stark}(high) = \hbar\omega_{eg} + \frac{\hbar\lambda^2}{(\omega_{eg} - \omega_{ph})} \dots\dots\dots (5)$$

It is easy to prove that, in the subspace containing the ground matter-state $|G, 0\rangle$ and the one-photon dressed down Floquet state $|Ex, -1\rangle$, the low-energy eigen state carrying the matter property is:

$$E^{down}_{Stark}(low) = -\frac{\hbar\lambda^2}{(\omega_{eg} - \omega_{ph})} \dots\dots\dots (6)$$

Thus, the final optical Stark shift that can be observed from the transition between the high-energy and low-energy eigen states is:

$$\delta E_{Stark} = E^{up}_{Stark}(high) - E^{down}_{Stark}(low) - \hbar\omega_{eg} = \frac{2\hbar\lambda^2}{(\omega_{eg} - \omega_{ph})} \dots\dots (7)$$

For the counterrotating part that corresponding to the Bloch-Siegert shift, the Hamiltonian can be written as:

$$\hat{H}_{Bloch-Siegert} = \hbar\omega_{eg}\hat{c}^\dagger\hat{c}^- + \hbar\omega_{ph}\hat{a}^\dagger\hat{a}^- + \hbar\lambda(\hat{c}^\dagger\hat{a}^\dagger + \hat{c}^-\hat{a}^-) \dots\dots\dots (8).$$

In the subspace containing the excited matter-state $|Ex, 0\rangle$ and the one-photon dressed down Floquet state $|G, -1\rangle$, (8) becomes:

$$\hat{H}^{down}_{Bloch-Siegert} = \hbar\omega_{eg}\begin{bmatrix} 1 & 0 \\ 0 & 0 \end{bmatrix} + \hbar\omega_{ph}\begin{bmatrix} 0 & 0 \\ 0 & -1 \end{bmatrix} + \hbar\lambda\begin{bmatrix} 0 & 1 \\ 1 & 0 \end{bmatrix} \dots\dots\dots (9)$$

The diagonalized eigen energy is

$$E^{down}_{Bloch-Siegert} = \frac{\hbar}{2}(\omega_{eg} - \omega_{ph}) \pm \frac{\hbar}{2}\sqrt{(\omega_{eg} + \omega_{ph})^2 + 4\lambda^2} \ \dots\dots (10)$$

since $\lambda \ll \omega_{eg} + \omega_{ph}$, (10) reduces to:

$$E^{down}_{Bloch-Siegert} = \frac{\hbar}{2}(\omega_{eg} - \omega_{ph}) \pm \frac{\hbar}{2}(\omega_{eg} + \omega_{ph}) \pm \frac{\hbar\lambda^2}{(\omega_{eg}+\omega_{ph})} \ \dots\dots (11)$$

The high-energy level carrying the matter property is:

$$E^{down}_{Bloch-Siegert}(high) = \hbar\omega_{eg} + \frac{\hbar\lambda^2}{(\omega_{eg}+\omega_{ph})} \ \dots\dots\dots\dots\dots\dots (12)$$

Similarly, in the subspace containing the ground matter-state $|G, 0\rangle$ and the one-photon dressed up Floquet state $|Ex, +1\rangle$, the low-energy level carrying the matter property is:

$$E_{Bloch-Siegert}^{up}(low) = -\frac{\hbar\lambda^2}{(\omega_{eg} + \omega_{ph})} \ldots \ldots \ldots (13)$$

Thus, the observed Bloch-Siegert shift is

$$\delta E_{\text{Bloch-Siegert}} = E_{\text{Bloch-Siegert}}^{down}(high) - E_{\text{Bloch-Siegert}}^{up}(low) - \hbar\omega_{eg}$$

$$= \frac{2\hbar\lambda^2}{(\omega_{eg} + \omega_{ph})} \ldots \ldots (14)$$

The ratio of the observed Bloch-Siegert shift to that of the optical Stark shift is $\frac{\omega_{eg} - \omega_{ph}}{\omega_{eg} + \omega_{ph}}$.